\documentclass[aps,prx,twocolumn,superscriptaddress,showpacs,longbibliography]{revtex4-2}

\usepackage{amsmath}
\usepackage{orcidlink}
\usepackage{graphicx}
\usepackage{amsfonts}
\usepackage{verbatim}
\usepackage{amssymb}
\usepackage{enumerate} 
\usepackage{enumitem} 
\usepackage{color}
\usepackage{dsfont}
\usepackage{amsmath} 
\usepackage{hyperref}
\usepackage{physics}
\hypersetup{colorlinks = true, urlcolor = blue, linkcolor = blue, citecolor = blue}

\newcommand{\bpm}{\begin{pmatrix}}
\newcommand{\epm}{\end{pmatrix}}

\newcommand{\nn}{\nonumber \\} 

\newcommand{\dg}{^{\dagger}}

\begin{document}

\title{Braiding Majoranas in a linear quantum dot-superconductor array: Mitigating the errors from Coulomb repulsion and residual tunneling}

\author{Sebastian Miles \orcidlink{0009-0005-6425-8072}}
\email{s.miles@tudelft.nl}
\affiliation{QuTech and Kavli Institute of Nanoscience, Delft University of Technology, Delft 2600 GA, The Netherlands}

\author{Francesco Zatelli \orcidlink{0000-0002-4250-6566}}
\affiliation{QuTech and Kavli Institute of Nanoscience, Delft University of Technology, Delft 2600 GA, The Netherlands}

\author{A. Mert Bozkurt \orcidlink{0000-0003-0593-6062}}
\affiliation{QuTech and Kavli Institute of Nanoscience, Delft University of Technology, Delft 2600 GA, The Netherlands}

\author{Michael Wimmer \orcidlink{0000-0001-6654-2310}}
\affiliation{QuTech and Kavli Institute of Nanoscience, Delft University of Technology, Delft 2600 GA, The Netherlands}

\author{Chun-Xiao Liu \orcidlink{0000-0002-4071-9058}}
\email{chunxiaoliu62@gmail.com}
\affiliation{QuTech and Kavli Institute of Nanoscience, Delft University of Technology, Delft 2600 GA, The Netherlands}

\date{\today}

\begin{abstract}
Exchanging the positions of two non-Abelian anyons transforms between many-body wavefunctions within a degenerate ground-state manifold. This behavior is fundamentally distinct from fermions, bosons and Abelian anyons.
Recently, quantum dot-superconductor arrays have emerged as a promising platform for creating topological Kitaev chains that can host non-Abelian Majorana zero modes.
In this work, we propose a minimal braiding setup in a linear array of quantum dots consisting of two minimal Kitaev chains coupled through an ancillary, normal quantum dot. We focus on the physical effects that are peculiar to quantum dot devices, such as interdot Coulomb repulsion and residual single electron tunneling.
We find that the errors caused by either of these effects can be efficiently mitigated by optimal control of the ancillary quantum dot that mediates the exchange of the non-Abelian anyons.
Moreover, we propose experimentally accessible methods to find this optimal operating regime and predict signatures of a successful Majorana braiding experiment.
\end{abstract}

\maketitle

\section{Introduction}

The exchange statistics of identical particles is a central concept in quantum mechanics.
It allows for classifying elementary particles (e.g., electrons and photons) into fermions and bosons.
In two-dimensional spaces, there exist more exotic particles, e.g., non-Abelian anyons~\cite{Moore1991Nonabelions, Nayak2008Non-Abelian}.
By exchanging the positions of two such anyons, referred to as a braid operation, the many-body wavefunction transforms into a different one in the degenerate ground-state manifold.
Thus, applying the same set of braid operations in a different order results in different unitary evolutions of the system.
In addition, non-Abelian anyons are regarded as the building blocks of topological quantum computation, where qubit information is encoded in a pair of anyons, and quantum gates are implemented by anyonic braiding~\cite{Nayak2008Non-Abelian,DasSarma2015Majorana}.
Ideally, this protocol is intrinsically fault-tolerant, because both storage and processing of the quantum information are immune to local perturbations due to the topological protection.
Therefore, demonstrating non-Abelian exchange statistics is of great importance to fundamental physics as well as to topological quantum computation.

Majorana zero modes, which are Ising anyons, are the simplest example of non-Abelian anyons~\cite{Alicea2012New,Leijnse2012Introduction,Beenakker2013Search,Stanescu2013Majorana,Jiang2013Non,DasSarma2015Majorana,Elliott2015Colloquium,Sato2017Topological,Aguado2017Majorana,Lutchyn2018Majorana,Zhang2019Next,Prada2020From}. 
They can appear at the defects of a topological superconductor in the form of a mid-gap quasiparticle excitation~\cite{Fu2008Superconducting,Sau2010Generic,Lutchyn2010Majorana,Oreg2010Helical}. 
In particular, it was proposed that topological Kitaev chains and Majorana zero modes can be engineered in a quantum-dot-superconductor array using a bottom-up approach~\cite{Sau2012Realizing}. 
An advantage of this proposal is the intrinsic robustness against the effect of disorder that is ubiquitous in mesoscopic systems~\cite{Liu2012Zero,Mi2014X,Pan2020Physical}.
Moreover, by controlling the relative strengths of normal and superconducting couplings between neighboring quantum dots~\cite{Liu2022Tunable,Bordin2023Tunable}, it is even possible to create Majoranas in the short-chain limit~\cite{Leijnse2012Parity}, albeit lacking true topological protection in this case.
Based on these proposals, significant experimental progress has been achieved recently in realizing short Kitaev chains in two-~\cite{Dvir2023Realization,tenHaaf2024Twosite,Zatelli2024Robust} and three-quantum-dot chains~\cite{Bordin2024Crossed,Bordin2024Signatures,tenHaaf2024Edge}, supported by tunnel spectroscopy evidence of Majorana zero modes at finely tuned sweet spots.
This opened up a new research field for Majorana physics and topological superconductivity~\cite{Liu2022Tunable,Tsintzis2022Creating,Liu2024Enhancing,Souto2023Probing,Koch2023Adversarial,Pandey2023Majorana,Bozkurt2024Interaction,Miles2024Kitaev,Luna2024Flux,Liu2024Protocol,vanDriel2024Cross,Liu2024Coupling,Ezawa2024Even_odd,Samuelson2024Minimal,Benestad2024Machine,Svensson2024Quantum,Souto2024Majorana,Pino2024Minimal,Souto2024Subgap,Nitsch2024Poor,Luethi2024From,Luethi2024Fate,Momez2024High,Pandey2024Crystalline,Alvarado2024Interplay}.
It also provides a new and promising platform to demonstrate the non-Abelian character of the exchange statistics~\cite{Liu2023Fusion,Boross2024Braiding,Pandey2024Nontrivial,Tsintzis2024Majorana}, which has been elusive for decades.

In quasi-one-dimensional systems, braid operations can also be implemented by cyclic tuning of the pairwise Majorana couplings in a trijunction~\cite{Sau2011Controlling,venHeck2012Coulomb,Karzig2015Shortcuts,Hell2016Time}, or by a sequence of measurement on the fermion parity in Majorana pairs~\cite{Bonderson2008Measurement,Plugge2017Majorana,Karzig2017Scalable}, both of which are mathematically equivalent to physically moving Majoranas  in a $T$-junction~\cite{Alicea2011NonAbelian}.
Furthermore, it was shown that the setup of trijunction braiding can be further simplified, where the role of a vertical topological superconductor branch can be replaced by a quantum dot~\cite{Liu2021Minimal, xu2023dynamics, Clarke2017Probability}.
However, it is a critical open question whether a braid protocol proposed for Majorana nanowires remains valid in the quantum dot setups with strong interactions.
For example, it has been recently shown that strong interdot Coulomb interaction can prevent the extraction of Majorana quality measures~\cite{Souto2023Probing}, and that Coulomb interaction within a Kitaev chain can be detrimental to the protection of Majorana zero modes or qubits~\cite{Leijnse2012Parity,Souto2024Majorana,Nitsch2024Poor}.

In the current work, we generalize the minimal braid protocol to engineered Kitaev chains, focusing on the physical effects that are peculiar to quantum dot devices, e.g., strong interdot Coulomb repulsion and residual single-electron tunneling.
Surprisingly, we find that the detrimental errors caused by both effects can be efficiently mitigated by optimal control of the ancillary quantum dot.
Moreover, we propose experimentally accessible methods to find this optimal operating regime and predict signatures of a successful Majorana braiding experiment.

\begin{figure}
    \centering
    \includegraphics[width=\linewidth]{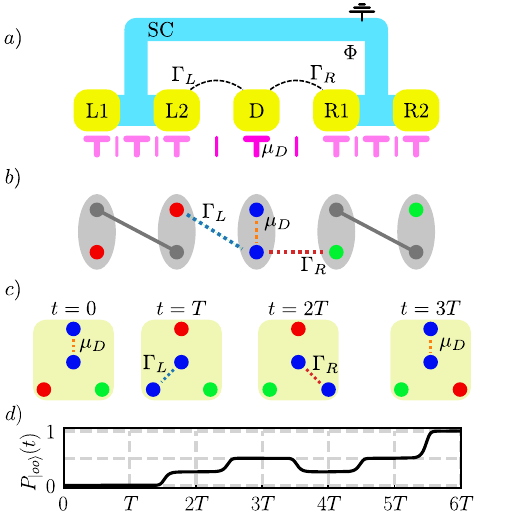}
    \caption{(a) Schematic of the minimal setup required for braiding in a linear array of quantum dots. Yellow squares are normal quantum dots, blue regions are superconducting leads mediating normal and Andreev tunneling. Purples lines are electrostatic gates to control the parameters.
    (b) Majorana representation of the Hamiltonian: The grey ovals with filled circles represent the Majorana operators $\gamma_{a, A/B}$ for dot $a$, lines represent effective couplings.  By tuning its chemical potential, the ancillary dot D supplies two Majoranas forming a virtual trijunction together with dots $L2$ and $R1$.
    (c) Schematic of the single Majorana exchange protocol. A full braid is implemented by varying $\mu_D$, $\Gamma_{L}$, and $\Gamma_R$ in sequence twice. $\mu_D, \Gamma_L$, and $\Gamma_R$ can be experimentally controlled via three electrostatic gates (dark purple).
    (d) Occupation probability of the $|oo\rangle$ state depending on $t$ over a full braiding operation. At times $3T$ and $6T$ the protocol implements exchange and full braid of the two Majoranas neighboring the ancillary dot respectively. The line highlights the change in parity the system undergoes during the protocol.
}
    \label{fig:schematic}
\end{figure}

\section{Setup and Model Hamiltonian}\label{sec:model}
The minimal braiding setup in a linear array of quantum dots consists of two copies of a two-site Kitaev chain connected by an ancillary quantum dot in the middle.
A schematic of this setup is shown in Fig.~\ref{fig:schematic}(a).
The model Hamiltonian is
\begin{align} 
& H = H_L + H_R + H_D + H_{\rm{tunnel}} + H_{\rm{Coulomb}}, \nn
& H_{a}= \sum_{i=1,2} \mu_{ai}n_{ai} + t_{a}c\dg_{a2} c_{a1} + \Delta_{a}c_{a2} c_{a1} + h.c., \nn
& H_D = \mu_D n_D, \nn
& H_{\rm{tunnel}} = \Gamma_L c\dg_D c_{L2} + \Gamma_R e^{i\varphi/2} c\dg_D c_{R1} + h.c., \nn
& H_{\rm{Coulomb}} = U_L n_D n_{L2} + U_R n_D n_{R1}.
\label{eq:H_total}
\end{align}
Here $H_{a}$ with $a=L/R$ are the Kitaev chain Hamiltonians, $c_{ai}$ and $n_{ai} = c\dg_{ai}c_{ai}$ are the annihilation and number operators of the dot orbitals, $\mu_{ai}$ is the orbital energy, and $t_a$ and $\Delta_a$ are the normal and Andreev tunnelings.
$H_D$ is the Hamiltonian for the ancillary quantum dot.
Here we assume that both the magnetic-field-induced Zeeman energy and the level spacing are large, such that all quantum dots are in the spinless regime.
The effect of onsite Coulomb interaction can thus be safely neglected.
We note that our spinless model is a very good approximation of the more microscopic quantum dot-superconductor system in the strong Zeeman regime ($E_{Z}$ in dots being much larger than $t_a,\Delta_a$), thus capturing all the key features of Majorana braiding.
On the other hand, including the spin degrees of freedom can cause minor overlap of Majorana zero modes that may cause decoherence, the effect of which will be partially studied in Sec.~\ref{sec:dephasing}.
$H_{\rm{tunnel}}$ describes single electron transfer between the end of the Kitaev chain and the ancillary quantum dot, with $\Gamma_{L/R}$ being the tunneling amplitudes.
$\varphi$ is the phase difference between the two superconducting leads, which can be controlled by the magnetic flux $\Phi$ through the loop. 
Here we choose a gauge such that $t_a,\Delta_a >0$.
$H_{\rm{Coulomb}}$ describes the interdot Coulomb interaction between the inner dots of the Kitaev chains and the ancillary dot.
We neglect Coulomb interaction between two dots of the same chain due to the strong screening effect of the grounded superconductor.

\section{Minimal Majorana braiding in a quantum dot chain} \label{sec:ideal_protocol}

\subsection{Effective trijunction in the Majorana representation}
One of the key results of Refs.~\onlinecite{Liu2021Minimal, xu2023dynamics} was that a quantum dot forming a junction between two Majorana bound states behaves as an effective tri-junction at a phase difference of $\pi$.
Here, we briefly show how this argument applies to the quantum dot chain.

To this end, we rewrite the Hamiltonian ~\eqref{eq:H_total} in the Majorana basis.
For each dot, we transform the fermionic operators into Majorana operators as
\begin{align}
    c_{a} = (\gamma_{aA} + i\gamma_{aB})/2, \quad c\dg_{a} = (\gamma_{aA} - i\gamma_{aB})/2.
\end{align}
At the sweet spot of the Kitaev chain we have
\begin{align}
    H_{L} +  H_{R} = i\Delta_L \gamma_{L2A}\gamma_{L1B} + i\Delta_R \gamma_{R2A}\gamma_{R1B},
\end{align}
with unpaired Majoranas $\gamma_{L1A}, \gamma_{L2B}, \gamma_{R1A}, \gamma_{R2B}$.
On the other hand, the Hamiltonian for the ancillary dot reads 
\begin{align}
    H_D = i\frac{\mu_D}{2} \gamma_{DA}\gamma_{DB}.  
\end{align}
The left tunneling Hamiltonian can be rewritten as
\begin{align}
    H_{tunn,L} = i\frac{\Gamma_L}{2} (\gamma_{L2A}\gamma_{DB} - \gamma_{L2B}\gamma_{DA}) \approx -i\frac{\Gamma_L}{2} \gamma_{L2B}\gamma_{DA}
\end{align}
where the approximation is to project away the coupling to the high-energy Majorana when $\Delta_L \gg \Gamma_L$.
For the right tunneling Hamiltonian, at $\varphi=\pi$, we have equally
\begin{align}
    H_{tunn,R} = i\frac{\Gamma_R}{2} (\gamma_{R1A}\gamma_{DA} + \gamma_{R1B}\gamma_{DB}) \approx i\frac{\Gamma_R}{2} \gamma_{R1A}\gamma_{DA}.
\end{align}
Thus the effective Hamiltonian is 
\begin{align}
    H_{\text{eff}} = i\frac{\mu_D}{2} \gamma_{DA}\gamma_{DB} -i\frac{\Gamma_L}{2} \gamma_{L2B}\gamma_{DA}- i\frac{\Gamma_R}{2} \gamma_{R1A}\gamma_{DA},
\end{align}
and thus equivalent to a Majorana trijunction \cite{venHeck2012Coulomb}, and schematically shown in Fig.~\ref{fig:schematic}(b).
Here, the dot energy $\mu_D$ plays effectively the role of a Majorana coupling.

\subsection{Braiding in the ideal case} \label{sec:id_braid}
We first consider an ideal scenario for Majorana braiding.
Assuming no interdot Coulomb interaction ($U_L=U_R=0$), two finely tuned Kitaev chains ($\mu_{ai}=0$ and $t_a=\Delta_a$) can host four zero-energy Majoranas 
\begin{align}
    &\gamma_1 = \gamma_{L1A}, \gamma_2 = \gamma_{L2B}, \gamma_3 = \gamma_{R1A}, \gamma_4 = \gamma_{R2B},
    \label{eq:4_MZM}
\end{align}
localized on four different quantum dots.
They form the degenerate ground-state manifold.
As shown above, when the phase condition $\varphi=\pi$ is satisfied, Majoranas $\gamma_2$ and $\gamma_3$ together with the ancillary quantum dot, form an effective trijunction, with the coupling strengths being $\Gamma_{L}, \Gamma_{R}$ and $\mu_D$, respectively.
Starting from uncoupled Majoranas with $\Gamma_L=\Gamma_R=0$ and $\mu_D>0$, we perform a sequence of three operations, adapting the protocol of Ref.~\cite{venHeck2012Coulomb}:

\begin{enumerate}[label=\arabic*., itemsep=5pt]
    \item turn off $\mu_D$ while turn on $\Gamma_L$, \quad $0 < t \leq T$
    \item turn off $\Gamma_L$ while turn on $\Gamma_R$, \quad $T < t \leq 2T$ 
    \item turn off $\Gamma_R$ while turn on $\mu_D$ to its original value, \quad $2T < t \leq 3T$.
\end{enumerate}

The effect is to exchange the positions of $\gamma_2$ and $\gamma_3$ as shown in Fig.~\ref{fig:schematic}(c). The action of the braid protocol is described by the operator

\begin{align}
    B = U(3T) = \exp \left \{ \frac{\pi}{4} \gamma_2 \gamma_3 \right \}.
    \label{eq:Braid}
\end{align}

Here we assume that all the operations are performed with perfect precision in the adiabatic limit and without any noise from the environment.
The effect of the braiding operation becomes apparent when tracking the time evolution of some initial state in the ground-state manifold through the time evolution.
Due to fermion parity conservation in Eq.~\eqref{eq:H_total}, we can focus on the subspace with total even parity without losing generality.
When $\Gamma_L=\Gamma_R=0$ and $\mu_D>0$, the ground states are doubly degenerate with 
\begin{align}
    & \ket{ee} \equiv \frac{1}{2} \left( \ket{00}_L - \ket{11}_L  \right) \otimes \left( \ket{00}_R - \ket{11}_R  \right) \otimes \ket{0}_D, \nn
    & \ket{oo} \equiv \frac{1}{2} \left( \ket{10}_L - \ket{01}_L  \right) \otimes \left( \ket{10}_R - \ket{01}_R  \right) \otimes \ket{0}_D,
    \label{eq:ee_oo}
\end{align}
where the basis states are defined as $\ket{n_{L1},n_{L2}} \otimes \ket{n_{R1},n_{R2}} \otimes \ket{n_D}$.
If the system is initialized as an even-even state
\begin{align}
    \ket{\psi(0)} = \ket{ee},
    \label{eq:psi_0}
\end{align}
it will evolve into 
\begin{align}
    \ket{\psi(3T)} = B \ket{\psi(0)} = (\ket{ee} - i \ket{oo})/\sqrt{2},
    \label{eq:psi_3T}
\end{align}
after performing the braid operation once.
By repeating the same braid operation, although Majoranas $\gamma_{2}$ and $\gamma_3$ return to their original positions, the system becomes 
\begin{align}
    \ket{\psi(6T)} = B^2 \ket{\psi(0)} = \ket{oo},
    \label{eq:psi_6T}
\end{align}
which is orthorgonal to the inital state.
Equations~\eqref{eq:psi_3T} and~\eqref{eq:psi_6T} are regarded as the signatures of non-Abelian statistics of Majorana anyons.
However, an experimental demonstration of Eq.~\eqref{eq:psi_3T} would be challenging.
What can be measured are probabilities $P_{|ee\rangle}(3T) = \abs{\langle ee | \psi(3T)\rangle}^2$ and $P_{|oo\rangle}(3T) = \abs{\langle oo | \psi(3T)\rangle}^2$, which do not contain the crucial information of the relative phase ($-i$) between the two basis states. 
Moreover, even a detection of $P_{|ee\rangle}(3T)=P_{|oo\rangle}(3T)=1/2$, which is consistent with Eq.~\eqref{eq:psi_3T}, cannot exclude the possibility of a completely decohered state with an uniform probability distribution.
In contrast, measuring the outcome of a double braid operation in Eq.~\eqref{eq:psi_6T}, which yields $P_{|ee\rangle}(6T)=0$ and $P_{|oo\rangle}(6T)=1$, will be transparent to interpret and thus more convincing. 
Therefore, in the rest of the work, we will focus on the double-braid protocol, which takes six steps of operations and a total time of $6T$, unless stated otherwise.  
A complete overview of the time-dependence of $P_{|oo\rangle}(t)$ is schematically shown in Fig.~\ref{fig:schematic} (d).

\subsection{Braiding in the imperfect case} \label{sec:imp_braid}

A real system will deviate from this ideal case.
For example, inter-dot Coulomb repulsion may lead to additional splittings, or some residual couplings between quantum dots may remain.
Additionally, the "leg" of the effective trijunction formed by the middle quantum dot is not protected.
Hence, noise in $\mu_D$ can be expected to have a significant impact.
Moreover, the phase-difference may deviate from the ideal value of $\pi$.

In the remainder of the paper, we will study the effects of these imperfections, and how to mitigate them.
To this end, we will use the full Hamiltonian \eqref{eq:H_total} with time-dependent parameters $\Gamma_{L/R}(t)$ and $\mu_D(t)$.
In particular, their time dependence profiles are chosen to be a sufficiently smooth hypertangent function, in order to minimize the additional diabatic errors caused~\cite{Knapp2016Nature}. 
We then compute $\psi(6T)$ by solving the time-dependent Schr\"odinger equation.
For more details on the simulations, we refer the reader to App.~\ref{sec:app_numerics}.

To characterize the faithfulness of the protocol, we calculate the infidelity
\begin{align} \label{eq:fidelity_def}
    1 - F \equiv 1 - \abs{\expval{oo | \psi(6T)}}^2 = 1 - P_{|oo\rangle}(6T),
\end{align}
where $\ket{\psi(6T)}$ is the final state after time evolution through a double braid protocol and $\ket{oo}$ is the analytical target state, respectively.
Note that the infidelity can be obtained experimentally from readout measurement on $P_{|ee\rangle}(6T)$ and $P_{|oo\rangle}(6T)$.

In our simulations, unless stated otherwise, we choose the system parameters to be $t_L = \Delta_L = t_R = \Delta_R=\Delta=5\Gamma_0, \mu_{L1} = \mu_{L2}=\mu_{R1} = \mu_{R2}=0$ to satisfy the sweet-spot condition, and  $\varphi=\pi$ for the phase condition.
Here, $\Gamma_0$ is the maximal strength of single electron tunneling setting the energy and time scale of the braiding process.
We make sure that the time evolution satisfies the adiabatic limit, i.e. $T \gg h/\Gamma_0$, and assume no environmental noise or quasiparticle poisoning.

\begin{figure*}
    \centering
    \includegraphics[width=\linewidth]{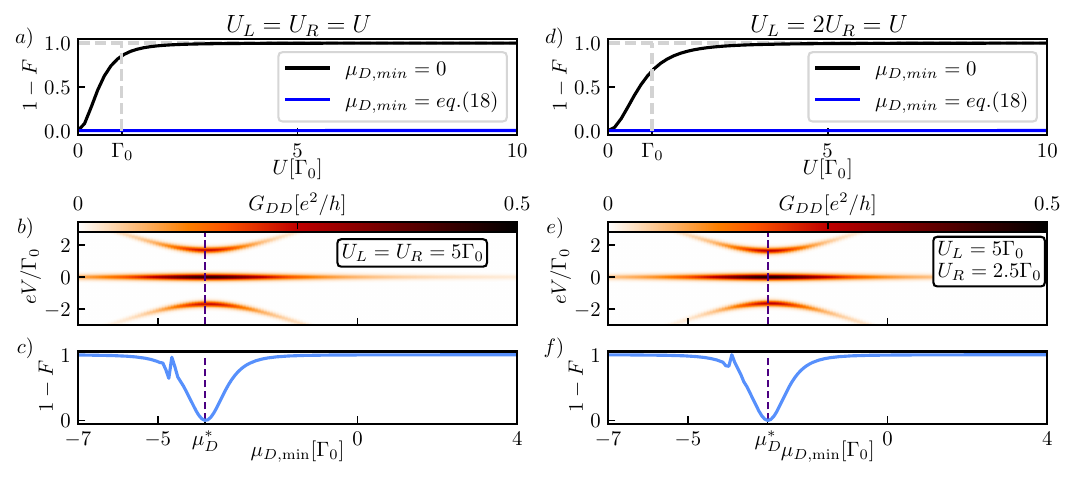}
    \caption{Effects of interdot Coulomb interaction between ancillary dot and adjacent Kitaev chain dots. (a) and (d) show the infidelity in dependence of symmetric and asymmetric Coulomb energy respectively. (b) and (e) show local conductance spectroscopy through the ancillary dot. Due to the interaction, the excitation minimum shifts in chemical potential to a lower value corresponding to Eq.~\eqref{eq:mu_Dstar}. Retuning $\mu_{D,\mathrm{min}}$ to this value corrects the adverse effect of the Coulomb interaction. This is supported by (c) and (f) showing the infidelity in dependence of $\mu_{D,\mathrm{min}}$. In line with the excitation minimum, the infidelity reduces to zero when $\mu_{D,\mathrm{min}}=\mu_D^*$. The discontinuity at $\mu_D^*-\Gamma_0$ is due to our choice of $\mu_{D,\mathrm{max}}-\mu_{D,\mathrm{min}}=\Gamma_0$ where the occupied state on the dot becomes resonant with the states in the Kitaev chains.  Measuring the traces as those presented in c) and f) experimentally can be considered a signature of Majorana braiding.}
    \label{fig:interdot_coulomb}
\end{figure*}

\section{Interdot Coulomb repulsion} \label{sec:interdot_coulomb}
We now consider the effect of interdot Coulomb repulsion on Majorana braiding.
Coulomb interaction is ubiquitous for quantum-dot-based devices, with the strength varying in a wide range of tens of $\mu$eV to as large as one meV~\cite{Hensgens2017Quantum, Hsiao2024Exciton}.
As described by $H_C$ in Eq.~\eqref{eq:H_total}, it is present among electrons on dots $L2, D$, and $R1$ due to the long-range nature of Coulomb interaction, while the interaction between dots within a Kitaev chain is strongly suppressed by the screening effect of the grounded superconductor. 

We begin by assuming that Coulomb interaction is present, but that the three time-varying parameters have equal variation magnitude and can be tuned perfectly to zero,
\begin{align}
    0 \leq \Gamma_{L}(t), \Gamma_{R}(t),\mu_{D}(t) \leq \Gamma_0,
    \label{eq:parameter_range}
\end{align}
before relaxing this assumption in later discussions.
As shown in Fig.~\ref{fig:interdot_coulomb} (a) ($U_L=U_R$) and (d) ($U_L\neq U_R$), interdot Coulomb energy has a very detrimental effect on braiding, with the infidelity quickly approaching one as $U \gtrsim \Gamma_0$.

To understand the physics behind this behavior, we focus on the first step of the braiding operation ($0 \leq t \leq T$).
Since the right Kitaev chain is decoupled in this process, we can work on a simpler Hamiltonian of $H_{LD} = H_{K,L} + H_{\text{tunn}, L} + H_{C, L} + H_{D}$.
Within the subspace of total even parity, it can be written as
\begin{align} \label{eq:h_ld}
    H^{(\text{even})}_{LD} =
    \begin{pmatrix}
        0 & -\frac{\Gamma_L}{2} & 0 & \frac{\Gamma_L}{2} \\
        -\frac{\Gamma_L}{2} & \mu_D + \frac{U_L}{2} & \frac{\Gamma_L}{2} & -\frac{U_L}{2} \\
        0 & \frac{\Gamma_L}{2} & 2\Delta_L & \frac{\Gamma_L}{2} \\
        \frac{\Gamma_L}{2} & -\frac{U_L}{2} & \frac{\Gamma_L}{2} & 2\Delta_L+\mu_D + \frac{U_L}{2}
    \end{pmatrix}
\end{align} 
where the basis is $\ket{e_L, 0_D}, \ket{o_L, 1_D}, \ket{e'_L, 0_D}, \ket{o'_L, 1_D}$ and primes indicate excited states.
Here we shift all states by $\Delta_L$ for simplicity of discussion, and the prime denotes the excited states in the Kitaev chain.
In the tunneling regime of $\Gamma_L \ll \Delta_L$, the low-energy effective Hamiltonian is
\begin{align}
    H^{(\text{even})}_{LD, \text{eff}} = 
    \begin{pmatrix}
        0 & -\frac{\Gamma_L}{2}(a+b) \\ 
        -\frac{\Gamma_L}{2}(a+b) & \mu_D + \frac{U_L}{2} + \Delta_L - \lambda
    \end{pmatrix}
    \label{eq:H_eff}
\end{align}
for arbitrary strength of $U_L$, and valid up to second order in $\Gamma_L$.
Here $\lambda= \sqrt{\Delta^2_L + (U_L/2)^2}$, and $a,b$ are positive numbers with $a^2 = 1-b^2 = \frac{1}{2} + \frac{\Delta_L}{2\lambda }$.
The low-energy basis states are $\ket{\psi_1} = \ket{e_L, 0_D}$, and $\ket{\psi_2} = a \ket{o_L, 1_D} + b \ket{o'_L, 1_D}$.
There are two major effects from the interdot Coulomb repulsion.
First, the instantaneous ground state of the total system now includes a component of the excited states $\ket{o'_L}$ in $\ket{\psi_2}$, compared to the idealized $\ket{\psi_2}=\ket{o_L, 1_D}$ for the case with $U_L=0$.
However, this change only takes place for intermediate states,  not affecting the form of the final wavefunction.
Thus no correction is needed for this effect.
Second, as shown in Eq.~\eqref{eq:H_eff}, the effective energy of the ancillary quantum dot is shifted: $\mu_D \to \mu_D + \frac{U_L}{2} + \Delta_L - \lambda$, which enhances the energy of $\ket{\psi_2}$.
In the strong Coulomb regime, $U\gg 1$, restricting $0 \leq \mu_D \leq \Gamma_0$ does not effectively take the dot down to resonance. 
As a result, the state $|\psi(t)\rangle$ would stay close to $\ket{ee}$ without moving any Majoranas, which explains the high infidelity in Fig.~\ref{fig:interdot_coulomb}(a).
Based on Eq.~\eqref{eq:H_eff}, one way to mitigate this detrimental error is to shift the dot energy as below
\begin{align}
    & \mu_{D,\text{min}} \leq \mu_D(t) \leq \mu_{D,\text{min}} + \Gamma_0, \nn
    & \mu_{D,\text{min}}= \mu^*_{D} = \sum_{a=L,R}\left( -\frac{U_a}{2} - \Delta_a + \sqrt{\Delta^2_a + (U_a/2)^2 } \right).
    \label{eq:mu_Dstar}
\end{align} 
Note that the dot energy shift now includes contributions from both Kitaev chains because the Coulomb potential is additive and we assume no coupling between the two Kitaev chains directly.
Applying Eq.~\eqref{eq:mu_Dstar} to the braid protocol and without changing any other conditions, we obtain the blue curve in Fig.~\ref{fig:interdot_coulomb} (a) for $U_L=U_R$ and (d) for $U_L\neq U_R$. 
It shows an excellent correction of the errors with $1-F \lesssim 10^{-3}$, validating our analysis and proposal.
Notably, since our treatment of Coulomb repulsion is nonperturbative in the interaction strength, the error mitigation applies to strong Coulomb case ($U > \Delta$) as well, provided the system stays in the tunneling regime $\Gamma_L \ll \Delta_L$.
Figure~\ref{fig:interdot_coulomb} (a) and (d) are the first main findings of this work, which positively indicates that it is possible to mitigate the detrimental effect of interdot Coulomb repulsion in Majorana braiding.

To find the value of $\mu^*_{D}$ in an actual device, we propose two experiments.
The first one is a local tunnel spectroscopy on the ancillary dot in the $(eV, \mu_{D})$ plane, as shown in Fig.~\ref{fig:interdot_coulomb}(b) and (e).
Here a normal lead is coupled to the ancillary dot, and a local conductance of $G_{DD}$ is obtained using the rate-equation approach~\cite{Tsintzis2022Creating,Liu2024Enhancing}.
Although the Majorana-induced zero-bias peak stays robust, the subgap peak from the first excited state varies with $\mu_D$, and reaches a minimum along the bias-voltage axis at $\mu_{D}=\mu^*_D$ [see Fig.~\ref{fig:interdot_coulomb}(b) and (e)], thus providing a way to find $\mu^*_D$ for the braid protocol.
Note that this does not add to the device complexity, when transport measurements are needed to fine-tune the Kitaev chains into their sweet spots.
Our second proposal is to measure the infidelity as a function of $\mu_{D,\text{min}}$.
As shown in Fig.~\ref{fig:interdot_coulomb}(c) and (f), the infidelity drops to nearly zero at the optimal value $\mu_{D,\text{min}}=\mu^*_D$, and then increases to one when $\mu_{D,\text{min}}$ is tuned away by $\sim \Gamma_0$.
The numerical simulations are consistent with our analytical results in Eq.~\eqref{eq:H_eff} and~\eqref{eq:mu_Dstar}.
Moreover, a measurement of Figs.~\ref{fig:interdot_coulomb}(c) or~\ref{fig:interdot_coulomb}(f) can be regarded as a signature of successful Majorana braiding. We note that the apparent discontinuity visible in Fig. \ref{fig:interdot_coulomb} (c) and (f) are a consequence of our initial choice $\max(\mu_D(t))=\Gamma_0$. When $\mu_{D,\mathrm{min}}\rightarrow \mu_D^*-\Gamma_0$ the occupied state of the ancillary dot is on resonance with the states on the Kitaev chain, interfering with the braiding. As this happens only for $\mu_{D,\mathrm{min}}<\mu_D^*$ and can be controlled by changing $\max(\mu_D(t))>\Gamma_0$, this feature can be disregarded for the effectiveness of the central result at $\mu_D^*$.

\section{Residual single electron tunneling} \label{sec:residual_tunneling}
\begin{figure*}
    \centering
    \includegraphics[width=\linewidth]{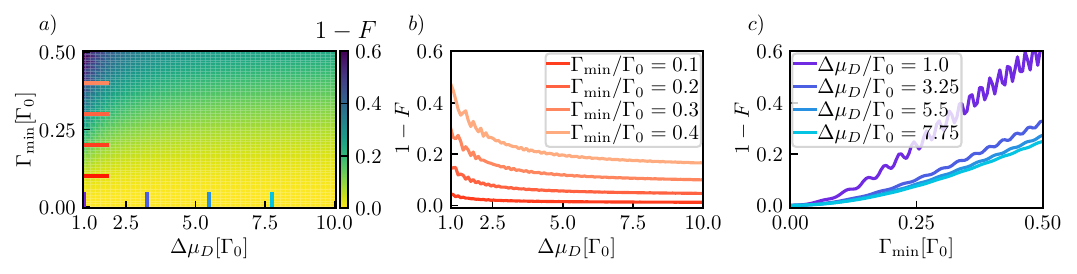}
    \caption{(a) Infidelity, $1-F$, in the $(\Delta \mu_D, \Gamma_{\text{min}})$-plane. (b) Infidelity as a function of $\Delta \mu_D$ for different cuts of $\Gamma_{\mathrm{min}}$ in (a). (c) Infidelity as a function of $\Gamma_{\mathrm{min}}$ for different cuts of $\Delta \mu_D$ in (a). }
    \label{fig:error_residual}
\end{figure*}

\begin{figure}
    \centering
    \includegraphics[width=\linewidth]{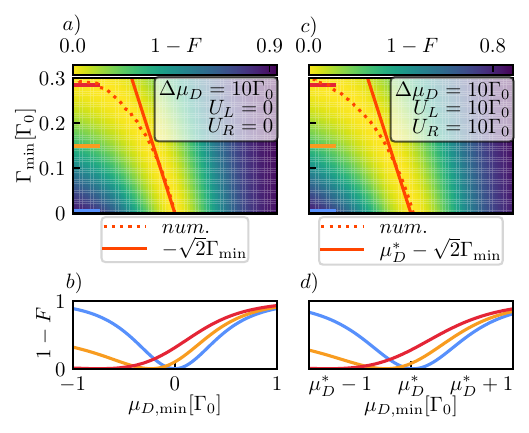}
    \caption{(a) and (c) Infidelity, $1-F$, in the $(\mu_{D,\text{min}}, \Gamma_{\text{min}})$ plane for $U_a=0$ and $U_a=10\Gamma_0$ respectively. For both, we choose $\Delta \mu_D=10\Gamma_0$. The dotted lines show the numerical minimum of the infidelity, solid lines correspond to the expectation of eqns.~\eqref{eq:mu_D_opt} and \eqref{eq:mu_Dstar}. (b) and (d) Infidelity as a function of $\mu_{D,\text{min}}$ for different cuts in $\Gamma_{\text{min}}$ through (a) and (c) showing that for increasingly negative values of $\mu_{D,\mathrm{min}}$ the infidelity vanishes regardless of residual tunnel coupling.
    Note that panels a) and b) share the same label for $x$ axes, and so do panels c) and d).}
    \label{fig:error_optimal}
\end{figure}

In semiconducting quantum dot devices, the single electron tunneling strength is controlled by electrostatic gates.
Although the strength can be varied deterministically, it is challenging to turn off the coupling completely, causing unwanted errors in qubit control~\cite{vanRiggelen2021Atwo,Arute2019Quantum}.
To study the effect of residual coupling we assume
\begin{alignat}{2} 
    \Gamma_{\text{min}} &\leq \Gamma_L(t), \Gamma_R(t) &&\leq \Gamma_0, \label{eq:param_residual_1} \\
    \mu_{D,\mathrm{min}} &\leq \quad \mu_D(t) &&\leq \mu_{D,\mathrm{min}}+\Delta \mu_D, \label{eq:param_residual_2}
\end{alignat}
where $\Gamma_{\text{min}}>0$ is the residual tunneling strength, and $\Delta \mu_D= \mu_{D,\text{max}}- \mu_{D,\text{min}}$ is the variation magnitude of the ancillary dot energy.
For the analytic considerations we set the interdot Coulomb energy at first to zero, unless stated otherwise (we relax this assumption in Fig.~\ref{fig:error_optimal} (c) and (d)).
Figure~\ref{fig:error_residual}(a) shows the numerically calculated infidelities in the $(\Delta \mu_D, \Gamma_{\text{min}})$ plane. The infidelity increases with the residual tunneling strength $\Gamma_{\text{min}}$ while it decreases with the dot variation magnitude $\Delta \mu_D$, see Figs.~\ref{fig:error_residual}(b) and~\ref{fig:error_residual}(c).
As it will be shown below, the infidelity is a joint consequence of two distinct error mechanisms, which we call leakage and geometrical leakage error.

The leakage error, on one hand, can be understood from a heuristic perturbation theory analysis:
At the initial time, treating the residual tunnelings $\Gamma_{a,\text{min}}$ as a perturbative effect, the state $\ket{ee}$ can leak into the excited states 
\begin{align}
    \ket{oe}=\frac{1}{2}(\ket{10}_L-\ket{01}_L)\otimes(\ket{00}_R-\ket{11}_R)\otimes \ket{1}_D \\
    \ket{eo}=\frac{1}{2}(\ket{00}_L-\ket{11}_L)\otimes(\ket{10}_R-\ket{01}_R)\otimes\ket{1}_D
\end{align}
with a characteristic amplitude of $\Gamma_{\text{min}}/\Delta \mu_D$.
Thus the leakage probability is
\begin{align}
    P_{\text{leak}} \propto \left( \frac{\Gamma_{\text{min}}}{\Delta \mu_D} \right)^2
    \label{eq:P_leak}
\end{align}
to leading order in residual tunneling strength.
We emphasize that such a leakage effect is time-independent, thus having a different physical origin as the dynamical effect of Landau-Zener transition.
The same argument can be made for any specific state expected at an intermediate step of the time evolution.
In light of Eq.~\eqref{eq:fidelity_def}, for a perfect double braid besides residual couplings of the Majoranas, it is therefore to expect that $1-F=1-P_{|oo\rangle} \propto  P_{\mathrm{leak}} \times const.$, with a constant proportional to $T$.
Indeed, as shown in Fig.~\ref{fig:error_residual}(c), the numerically calculated infidelity decays with a larger variation magnitude, consistent with Eq.~\eqref{eq:P_leak}.
As indicated by Eq.~\eqref{eq:P_leak}, leakage errors can be mitigated by an increase of the chemical potential variation on the ancillary dot, i.e.

\begin{align}
    \Delta \mu_D \gg \Gamma_0.
\end{align}

We note that an increase of $\Delta \mu_D$ comes at the expense of diabatic errors due to faster changes of $\mu_D(t)$. 
We do however expect the existence of a window in protocol time where the leakage error is strongly suppressed before the diabatic error becomes prominent.

The geometrical error, on the other hand, can be understood by calculating the unitary evolution operator, Eq.~\eqref{eq:Braid}, in the presence of residual couplings between the Majoranas.
This amounts to calculating the non-Abelian Berry phase of the cyclic variation of the Hamiltonian, which has a geometrical origin.
Performing a similar calculation to that of Ref.~\cite{venHeck2012Coulomb} generalized to asymmetric couplings (see App.~\ref{sec:berry_phase}), the unitary operator in Eq.~\eqref{eq:Braid} becomes
\begin{align} \label{eq:Braid_eps}
    U_{\text{res}}(3T) =  \exp \left \{ \left( \frac{\pi}{4} - \epsilon \right) \gamma_2 \gamma_3 \right \},
\end{align}
where $\epsilon$ denotes the deviation from perfect braiding.
The corresponding infidelity is 
\begin{align} \label{eq:geom_infidel}
    1-F = \sin^2(2\epsilon) \approx 4 \epsilon^2
\end{align}
when $\epsilon \ll 1$.
Calculating $\varepsilon$ in Eq.~\eqref{eq:Braid_eps} explicitly, we find
for $\Gamma_{L,\text{max}}=\Gamma_{R,\text{max}}=\mu_{D,\text{max}}=\Gamma_0$

\begin{align}
    \epsilon \approx \frac{1}{\sqrt{2}\Gamma_0}\left( \Gamma_{L,\text{min}} + \Gamma_{R,\text{min}} + \mu_{D,\text{min}}\right)
    \label{eq:geo_error_sym}
\end{align}

in leading order of the residual coupling of the Majoranas (see App.~\ref{sec:berry_phase}).
We note that Eq.~\eqref{eq:geo_error_sym} coincides with the result in Ref.~\cite{venHeck2012Coulomb} when $\Gamma_{L,\text{min}}=\Gamma_{R,\text{min}}=\mu_{D,\text{min}}$.
Assuming no residual coupling of the Majoranas on the ancillary dot, i.e. $\mu_{D,\mathrm{min}}\equiv 0$ in the absence of Coulomb repulsion, we find through Eq.~\eqref{eq:geom_infidel} that the infidelity scales quadratically with the residual tunnel coupling,

\begin{align}
    1-F \propto \left( \frac{\Gamma_{\text{min}}}{\Gamma_0} \right)^2.
\end{align} 

We can find an analytical correction to the geometrical error when excluding the presence of leakage errors, that is in the limit $\Gamma_{L,\text{max}}=\Gamma_{R,\text{max}}=\Gamma_0 \ll \mu_{D,\text{max}}$.
We then obtain
\begin{align}
    \epsilon \approx \frac{1}{\sqrt{2}\Gamma_0}\left( \Gamma_{L,\text{min}} + \Gamma_{R,\text{min}} + \sqrt{2}\mu_{D,\text{min}} \right),
    \label{eq:geo_error_asym}
\end{align}
to leading order in residual tunnelings and in $\Gamma_0/\mu_{D,\text{max}}$.
The key observation now is that, despite $\Gamma_{a,\mathrm{min}}/\Gamma_{a,\mathrm{max}}>0$ for any applied voltage to the tunnel gates, one can change the sign of $\mu_{D,\mathrm{min}}$ with respect to $\mu_{D,\mathrm{max}}$ by tuning the dot to chemical potentials below the resonance of dot with the Kitaev chains.  
In particular, the geometrical error vanishes up to the leading (quadratic) order in $\Gamma_{\text{min}}/\Gamma_0$ when one chooses
\begin{align}
    \mu_{D, \text{min}} = -\sqrt{2}\Gamma_{\text{min}}<0.
    \label{eq:mu_D_opt}
\end{align}

Figure \ref{fig:error_optimal} (a) shows the infidelity of the double braid protocol in the $(\mu_{D.\mathrm{min}}, \Gamma_{\mathrm{min}})$ plane for $\Delta \mu_D=10\Gamma_0$.
The dotted red line indicates the numerical minimum of the infidelity, $1-F$, for fixed $\Gamma_{\mathrm{min}}$ in dependence of $\mu_{D,\mathrm{min}}$ while the solid red line shows Eq.\eqref{eq:mu_D_opt}.
Indeed, the optimized $\mu_{D,\text{min}}$ take negative values as predicted. Moreover, the analytic result in Eq.~\eqref{eq:mu_D_opt} matches well with the numerical result in the weak residual tunneling regime .
We furthermore find that our analysis remains well valid even in the presence of Coulomb repulsion when additionally applying the correction suggested in Sec.~\ref{sec:interdot_coulomb}, i.e. $\mu_D \to \mu_D + \mu^*_D $ as visible in Fig.~\ref{fig:error_optimal} (c) and (d).

\section{Superconducting phase difference}
\label{sec:pi_phase}
\begin{figure}
    \centering
    \includegraphics[width=\linewidth]{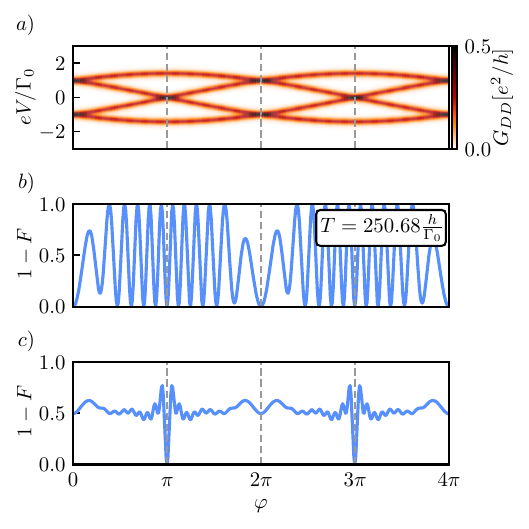}
    \caption{(a) Tunnel spectroscopy, $G_{DD}$, over the ancillary dot in the $(V, \varphi)$ plane. Only at odd integer multiples of $\pi$ the conductance indicates the necessary degeneracy at $V=0$. Additionally, the linear splitting of that degeneracy with phase indicates the lack of protection of the protocol against phase noise. (b) Infidelity, $1-F$, as a function of $\varphi$ for a single $T$. The oscillations indicate Rabi oscillations between the states in the ground-state manifold. (c) Infidelity of the double braid protocol averaged over multiple $T$. Since the outcome of the non-Abelian exchange does not depend on any specific choice of $T$, the perfect fidelities at odd interger multiples of $\pi$ persist while the Rabi oscillations present in (b) average away.}
    \label{fig:pi_phase}
\end{figure}

Satisfying the phase condition $\varphi=\pi$ is crucial for a successful braiding experiment.
The phase is controlled via the magnetic flux through the superconducting loop, i.e. $\varphi = 2\pi e\Phi/h + const.$, see Fig.~\ref{fig:schematic}(a).
Here we propose two experiments to find where $\varphi=\pi$, which are similar in spirit to those discussed in Sec.~\ref{sec:interdot_coulomb}.
The first one is a transport measurement.
Figure~\ref{fig:pi_phase}(a) shows the tunnel spectroscopy of $G_{DD}$ in the $(eV, \varphi)$ plane.
The signature of $\varphi=\pi$ is a zero-bias conductance peak, which is induced by the Majorana zero modes formed at the trijunction and splits linearly when the phase is away from $\pi$.
Our second proposed experiment is to measure the double-braid infidelity $1-F$ as a function of $\varphi$, as shown in Fig.~\ref{fig:pi_phase}(b).
Interestingly, in addition to $\varphi=\pi$, there are multiple other values of $\varphi$ also giving zeros of $1-F$.
These zeros are due to Rabi oscillations induced by the undesired ground-state degeneracy splitting, similar to the observations in Ref.~\cite{Liu2021Minimal,Pan2024Rabi}.
However, a fundamental distinction between them is that the outcome of non-Abelian braiding does not depend on the precise control of the protocol time as the dynamical effects such as Rabi oscillations.
Therefore, after averaging over different lengths of protocol time $T$, while keeping the adiabaticity constraint still satisfied, only the infidelity at $\varphi=\pi$ remains zero [see Fig.~\ref{fig:diabatic_dephasing}(c)], indicating the robustness of a geometrical braid operation.

\begin{figure}
    \centering
    \includegraphics[width=\linewidth]{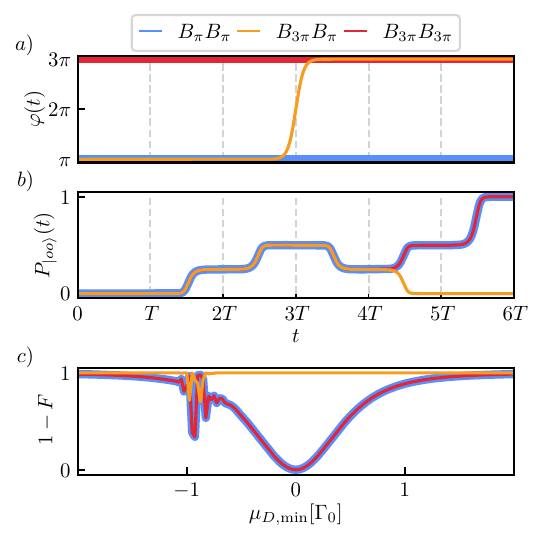}
    \caption{Time and dot chemical potential traces demonstrating the $4\pi$-Josephson effect of the system for the system specified in Sec.~\ref{sec:imp_braid} up to phase. We choose $\Delta \mu_D=1$ and neglect interdot Coulomb repulsion. (a) Superconducting phase $\varphi$ in dependence of time. To implement $B_\pi B_\pi$ and $B_{3\pi} B_{3\pi}$ we keep the phase fixed while for $B_{3\pi} B_\pi$ we adiabatically change the phase from $\pi \rightarrow 3\pi$ around $t= 3T$. (b) occupation probability of the $|oo\rangle$ state. Only for $B_{3\pi} B_{\pi}$, the initialized $|ee\rangle$ state returns into itself over the time evolution. (c) Infidelity, $1-F$, depdending on $\mu_{D,\mathrm{min}}$. Only the $B_{\pi}B_{\pi}$ and $B_{3\pi}B_{3\pi}$ show the transition from $|ee\rangle \rightarrow |oo\rangle$ predicted for non-Abelian exchange. For $B_{3\pi}B_{\pi}$ the initial state returns to itself identically.}
    \label{fig:b_protocols}
\end{figure}

In addition, we notice that both the conductance spectroscopy and the infidelities in Fig.~\ref{fig:pi_phase} are $2\pi$-periodic in $\varphi$, or equivalently $h/2e$-periodic in magnetic flux $\Phi$.
However, the tunneling $\Gamma_R$ in Eq.~\eqref{eq:H_total} at $\varphi=3\pi$ acquires a minus sign relative to $\varphi=\pi$, giving
\begin{align}
    B_{3\pi} = B^{-1}_{\pi},
    \label{eq:braid_3pi}
\end{align}
where $B_{\pi}$ is defined in Eq.~\eqref{eq:Braid}.
This is a consequence of the $4\pi$ Josephson effect due to fractionalized Majorana zero modes.
In particular, single braids $B_{\pi}$ and $B_{3\pi}$ give $(\ket{ee} \pm i\ket{oo})/\sqrt{2}$, respectively.
However, it is challenging to distinguish the different phases $\pm i$ here from a measurement of $P_{|ee\rangle}$ and $P_{|oo\rangle}$ only.
Thus we propose the following three-step experiment.
\begin{enumerate}[label=\arabic*., itemsep=5pt]
    \item apply $B_{\pi}$ on $\ket{ee}$ twice to obtain $\ket{oo}$
    \item apply $B_{3\pi}$ on $\ket{ee}$ twice to obtain $\ket{oo}$
    \item apply $B_{\pi}$ on $\ket{ee}$ once followed by another $B_{3\pi}$ to obtain $\ket{ee}$
\end{enumerate}
Here in each step the system should be initialized at $\ket{ee}$.
A successful implementation of the above experiments would manifest $4\pi$-periodicity in a fractional Josephson junction.
We demonstrate our proposal in Fig.~\ref{fig:b_protocols} for the system described in Sec.~\ref{sec:imp_braid} for $\Delta \mu_D=1$ and without Coulomb repulsion.
We see that, if the phase stays constant over the double braid, the Majoranas exchange as expected and the quantum state of the system changes. 
If, however, the phase is adiabatically changed for the second exchange the state returns to the initial state due to the $4\pi$-Josephson effect.

\section{Diabatic and dephasing effects}\label{sec:dephasing}

\begin{figure*}
    \centering
    \includegraphics[width=\linewidth]{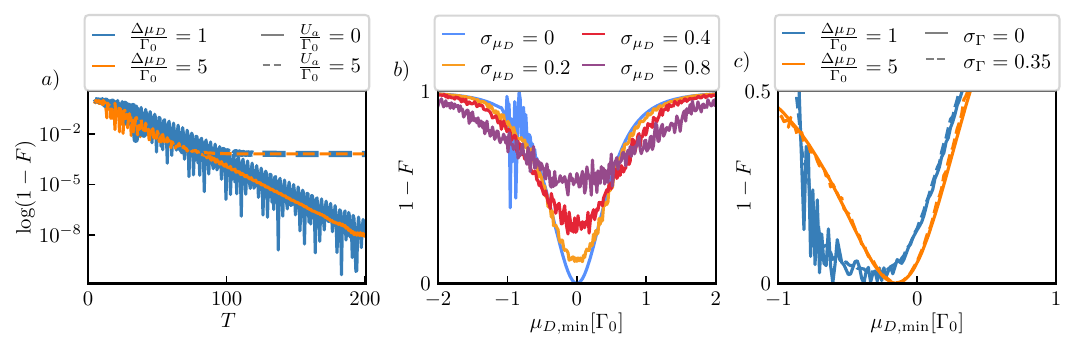}
    \caption{(a) Logarithm of the infidelity, $\log{1-F}$, in dependence of protocol step time $T$ showing oscillations due to leakage into excited states. The reduced fidelity at finite $U$ is due to perturbative corrections to $\mu_D^* ~ \Gamma_a$ not taken into account in Eq.~\eqref{eq:mu_Dstar}. (b) Ensemble averaged infidelity of the double braid in dependence of $\mu_{D,\mathrm{min}}$ for different variances of quasistatic noise on $\mu_D$ averaged over 100 realizations . (c) Ensemble averaged infidelity in dependence of $\mu_{D,\mathrm{min}}$ at fixed residual tunneling $\Gamma_{\mathrm{min}}=0$ over 100 noise values. Solid and dashed lines correspond to noise with a variance of $0$ and $0.35\Gamma_0$ on the tunnel coupling $\Gamma_a$ respectively.}
    \label{fig:diabatic_dephasing}
\end{figure*}

\label{sec:diabatic_dephasing}
We now consider the impact of diabatic and dephasing errors when executing the braid protocol.
Figure~\ref{fig:diabatic_dephasing} shows the infidelity as a function of the protocol step time $T$ at time $t=6T$ for different values of $\Delta \mu_D$ and $U$.
The infidelity decreases exponentially with the protocol time, consistent with the behavior of diabatic error in holonomy or anyonic braiding~\cite{Knapp2016Nature}.
Interestingly, by comparing the blue and orange lines in Fig.~\ref{fig:diabatic_dephasing}(a), we find that increasing $\Delta \mu_D$ decreases the diabatic error.
Physically, although the change of $\mu_D(t)$ becomes faster, the energy gap of the effective trijunction increases, which compensates the former effect as long no Landau-Zener transitions into the excited manifold are induced.
This means that we can suppress the leakage in Eq.~\eqref{eq:P_leak} without increasing the diabatic error.
Additionally, with finite Coulomb, the infidelity saturates at $1-F \approx 10^{-3}$ for $T\rightarrow \infty$, which is due to the higher-order corrections to $\mu_D^*$ from $\Gamma_{a}$ that are not included in leading-order result shown in Eq.~\eqref{eq:mu_Dstar}.

In semiconducting quantum dot devices, charge noise is the primary source of noise, which can be caused by charge impurities or gate voltage fluctuations~\cite{Hu2006Charge,Petersson2010Quantum, Dial2013Charge, Scarlino2022Insitu, Connors2022Charge, Throckmorton2022Crosstalk, Burkard2023Semiconductor, Paladino2014f}.
Since it is $1/f$ noise, which is dominated by the low-frequency component, we can model the noises using the quasi-static disorder approximation~\cite{Ithier2005Decoherence, Boross2022Dephasing}.
In particular, we focus on the effect of noise in the ancillary dot that does not exhibit any protection, whereas noisw in the short Kitaev chains could be mitigated by extending the chain length.
Moreover, the dephasing effects of noise within the Kitaev chains were studied in previous work in the context of imperfect Majorana polarizations~\cite{Tsintzis2024Majorana}.

We assume the noise on the ancillary dot chemical potential can be modeled as
\begin{align}
    \mu_D(t) \to \mu_D(t) + \delta \mu_D,,
\end{align}
where $\delta_{\mu_D}$ is a constant shift drawn from a normal distribution with width $\sigma_{\mu_D}$ and centered around zero for each execution of the protocol(see App. \ref{sec:app_numerics}).
To quantify the effect of the noise we perform an ensemble average over $100$ different noise values.
As shown in Fig.~\ref{fig:diabatic_dephasing}(b), the main effect of $\mu_D$-noise is to deteriorate the fidelity around $\mu_{D,\mathrm{min}}=\mu_D^*$ ($=0$ when $U_a=0$).
Since the width of the $1-F$ dip is of the order of $\Gamma_0$, a necessary condition for observing this signature is a characteristic disorder strength $\mu_{D, \text{dis}} \ll \Gamma_0$ to indicate the success of the braid.

In contrast to chemical potential noise, the modeling of the tunneling noise, $\Gamma$, differs due to its distinct dependence on the electrostatic potential.
In particular, the quantum dot energy has a linear dependence because it is capacitively coupled to the electrostatic potential nearby, while the electron transfer rate has an exponential dependence since it is determined by the transmission probability through a tunnel barrier.
Therefore, noise in $\Gamma_a$ is proportional to $\Gamma_a$ itself, i.e.
\begin{align}
    \Gamma_a(t) \to \left( 1 + \alpha \right)\Gamma_a(t),
    \label{eq:Gamma_noise}
\end{align}
Here $\alpha$ is a dimensionless coefficient which we draw from a normal distribution of width $\sigma_\Gamma$ and mean zero.
We use this simple model of noise on tunneling, as it avoids taking into account details of the tunnel barrier, but we expect it is sufficient to demonstrate the principal effect of noise in the tunnel barriers when ensuring $\sigma_\Gamma \ll 1$.
As shown in Fig.~\ref{fig:diabatic_dephasing}(c), although the $1-F$ dip is again lifted by the noise, its effect is much weaker to the effect of chemical potential noise as shown in Fig.~\ref{fig:diabatic_dephasing}(b).
This feature can be well understood using Eq.~\eqref{eq:Gamma_noise}.
When the tunneling $\Gamma_a$ is in the \emph{off}-state, the fluctuation is strongly suppressed due to the small residual tunneling amplitude.
On the other hand, the deviation of $\Gamma_a$ strength in the \emph{on}-state predominantly changes the energy gap of the effective trijunction, without greatly affecting the outcome braiding operation.
Thus, tunneling noise is less detrimental than noise on $\mu_D$ in this braid setup.

\begin{figure}
    \centering
    \includegraphics[width=\linewidth]{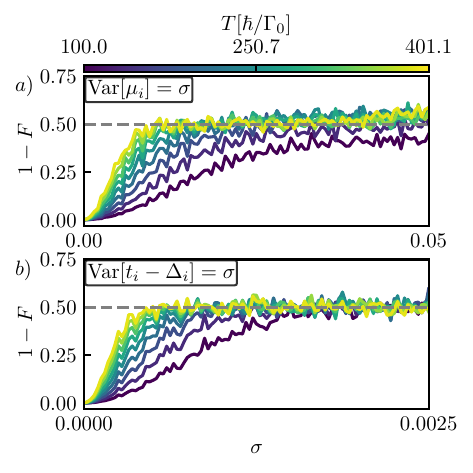}
    \caption{The dependence of infidelity on noises in chemical potential, a), and coupling strengths, b), in both Kitaev chains. The infidelity $1-F$ increases faster with the noise amplitude $\sigma$ for longer protocol time $T$.}
    \label{fig:pmm_noise}
\end{figure}

Finally, we allow for noise induced detuning on the Kitaev chains themself.
When detuned off the sweet spot, the groundstate degeneracy of a Kitaev chain splits.
This leads to oscillations between nearly degenerate ground states on a timescale $\hbar/\delta E$, where $\delta E$ is the ground-state energy splitting.
These oscillations deteriorate the fidelity of the final braid outcome via phase accumulation $\sim e^{i \delta E T/\hbar}$ 
We demonstrate this behavior in Fig.~\ref{fig:pmm_noise}.
On both Kitaev chains we allow for quasistatic noise on the indicated parameters, i.e. $\mu_i$ in a), $t_i-\Delta_i$ in b).
In Fig.~\ref{fig:pmm_noise} a), we let $\mu_i\rightarrow \mu_i+\delta\mu$, where $\delta\mu$ is drawn from a normal distribution with mean zero and variance $\sigma$.
Similarly, in Fig.~\ref{fig:pmm_noise} b), we let $t_i\rightarrow t_i+\delta t$, and draw the noise analogously.
Additionally, we vary the time $T$ to change the execution time of the protocol.
Fig.~\ref{fig:pmm_noise} implies a compromise between the protocol time and adiabaticity condition, and also shows the differing susceptibility to each error mechanism.
Given recent experimental results, we expect noises can be reduced by both enhancing the energy gap of the Kitaev chain, and reducing the chemical potential fluctuations by charge screening due to superconductor proximity effect~\cite{Zatelli2024Robust}.

\section{Discussion}
\label{sec:discussion}
It has been shown that braiding of non-Abelian anyons can take place only in two-dimensional space, which seemingly contradicts with the conclusions of the current work.
However, we emphasize that although the quantum dot-superconductor array has a linear structure at the first glance, it is quasi-one-dimensional in nature.
In particular, since $\varphi=\pi$ is a crucial requirement for a successful Majorana braid, a superconducting loop [see Fig.~\ref{fig:schematic}(a)] has to be formed along with a controllable magnetic flux $\Phi$, which extends the setup geometry to the second dimension. 
Moreover, we emphasize that the proposed braiding setup allows a ``minimal braiding" experiment in the sense that the outcome is \emph{not} topologically protected, but depends on fine-tuning of parameters, in particular $\mu_D$ and $\phi$.
In addition, both charge noises on the ancillary normal dot and noises within the Kitaev chains can cause decoherence in the final outcome.

We here discuss several relevant time scales of the braiding protocol. 
First, the protocol time should be sufficiently long in order to satisfy the adiabatic condition.
Although there is no common standard, here we choose a threshold diabatic error to be $10^{-2}$ for concrete discussions. 
According to Fig.~\ref{fig:diabatic_dephasing}(a), the protocol time needs to be larger than $\sim 300\hbar/\Gamma_0$, which correponds to a time scale of $\sim 20~$ns for a typical single electron tunneling strength of $\Gamma_0 \sim 10~\mu$eV.

Second, as discussed in Sec.~\ref{sec:diabatic_dephasing}, the dephasing effect from noises in $\mu_D$ and those within Kitaev chains should be sufficiently mitigated.
In particular, in order to experimentally observe Fig.~\ref{fig:interdot_coulomb}(c) or (f), which is regarded as one of the signatures of a successful Majorana braiding, the amplitude of the $\mu_D$ noise must be smaller than the characteristic single electron tunneling strength, i.e.,
\begin{align}
    \mu_{D,\text{dis}} \ll \Gamma_0,
\end{align}
as shown in Fig.~\ref{fig:diabatic_dephasing}(b).
Additionally, the energy splitting between the instantaneous ground states should also be much weaker than $\Gamma_0$ to avoid decoherence.
This can be achieved by either enhancing the excitation energy gap $\Delta_{L/R}$ of a Kitaev chain defined in Eq.~\eqref{eq:H_total}~\cite{Liu2024Enhancing,tenHaaf2024Twosite,Zatelli2024Robust} or extending the chain length~\cite{Liu2024Protocol,Bordin2024Signatures,tenHaaf2024Edge, Ezawa2024Even_odd}.
By contrast, based on our simulations and arguments, the noises in $\Gamma_{L/R}$ are less detrimental. 

The third time scale is the quasiparticle poisoning effect.
For example, a random incoming electron from outside the system can flip the total fermion parity, causing leakage errors that cannot be corrected.
The poisoning time is reported to be around $\sim 1~$ms in devices of InSb/InAs semiconductor nanowires proximitized by Al superconductors~\cite{Aghaee2024Interferometric}, which has a very similar nanostructure to the Kitaev chain devices~\cite{Dvir2023Realization,tenHaaf2024Twosite,Zatelli2024Robust,Bordin2024Crossed,Bordin2024Signatures,tenHaaf2024Edge}.
As long as the adiabatic condition is satisfied (e.g., $\sim 20~$ns for diabatic error $<10^{-2}$) within that time scale, the quasiparticle poisoning effect should not be a major concern in this braid experiment.
Finally we want to discuss some design aspects beyond our requirement of a minimal design.
Based on ideas presented in Ref.~\cite{Aghaee2025Distinct}, one extension is to introduce additional dots between chains and the auxilliary dot.
Tuning these buffer dots to be empty, away from resonance, one would change the direct coupling to a cotunneling.
While this would not fully remove the presence of residual tunneling, its strength might reduce.

Although the current work focuses on the even-parity subspace of the Hamiltonian, we expect that a similar analysis and correction scheme also applies to the odd-parity one, because the low-energy effective Hamiltonian has an analogous form as derived in Appendix~\ref{sec:app_coulomb}. 
On the other hand, system initialization and final outcome readout are also crucial for a successful braid experiment.
Here we propose to make use of the technique of quantum capacitance measurement which has been recently implemented in a single minimal Kitaev chain~\cite{vanLoo2025Single}.
When generalizing to the double Kitaev chain system as in this work, such a measurement on the even-parity subspace shows a much larger signal-noise-ratio~\cite{Liu2023Fusion}, and thus a braiding in the even-parity subspace would be much preferred.

\section{Summary}
\label{sec:summary}
In summary, we have investigated a minimal Majorana braiding protocol in quantum-dot-based Kitaev chains, focusing on the physical phenomena that are peculiar to quantum dot devices, e.g., interdot Coulomb repulsion and residual single electron tunneling.
We find that the detrimental errors from them can be efficiently mitigated by optimal control of the ancillary quantum dot via $\mu_{D,\text{min}}$ and $\mu_{D,\text{max}}$.
Furthermore, we propose a series of experiments to find this optimal operating regime and predict signatures of a successful braiding.
We also analyze the diabatic errors and dephasing effect from various types of noises.

\emph{Author contributions}
C.X.L., F.Z., and M.W. initiated the project. C.X.L. and S.M. designed the project. S.M. and C.X.L. performed the calculations with input from A.M.B.. C.X.L. supervised the project with input from M.W.. S.M., C.X.L, and M.W. wrote the manuscript with input from all authors. 

\emph{Acknowledgements}
We acknowledge useful discussions with Alberto Bordin, Bart Roovers, Florian Bennebroek Evertsz, and Juan Daniel Torres Luna about current experiments and their analysis.
This work was supported by a subsidy for top consortia for knowledge and innovation (TKI toeslag), by the Dutch Organization for Scientific Research (NWO) and Microsoft Station Q. S.M. acknowledges funding of NWO through OCENW.GROOT.2019.004. 

\emph{Data availability}
The code and the data that was generated for the plots are available in the repository of Ref. \cite{zenodo_repo}

\bibliography{references_CXL}

\onecolumngrid
\appendix

\section{Details of the numerical calculation} \label{sec:app_numerics}
In this appendix we expand on the numerical tools used to generate the figures of this work.
The associated repository containing all codes used for this work can be found in Ref. \cite{zenodo_repo}.

To simulate braiding in the setup described in Sec. \ref{sec:model}, we use the \textsc{QuTiP} \cite{johansson2012qutip} Python package to calculate the time evolution.
The perturbation theory results of Sec. \ref{sec:interdot_coulomb} and App. \ref{sec:app_coulomb} are generated with Pymablock \cite{Day2024Pymablock}.

To model the time dependent coupling between the different Majoranas (cf. Fig. \ref{fig:schematic} c)) we have to specify the profiles for $\mu_D, \Gamma_L$, and $\Gamma_R$.
We model each parameter by the time dependent function

\begin{align}
p(t, p_{min}, p_{max}, \sigma_p) &= (p_{max}-p_{min})\left(\frac{1}{2}+\tanh(\frac{\tilde{t}-\frac{T}{2}}{t_{ramp}})
-\tanh(\frac{\tilde{t}-\frac{3T}{2}}{t_{ramp}})+\tanh(\frac{\tilde{t}-\frac{7T}{2}}{t_{ramp}}) \right), \\
\tilde{t} &= (t+t_0)\mod 3T.
\end{align}

\begin{figure}
    \centering
    \includegraphics[width=0.5\linewidth]{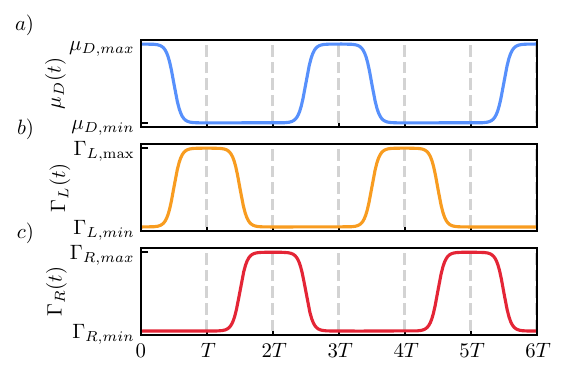}
    \caption{Parameters of the double braid protocol depending on time. (a) chemical potential $\mu_D(t)$ of the ancillary dot. (c) and (d), tunnel coupling of the ancillary dot to the left and right Kitaev chain respecitvely.}
    \label{fig:app_profiles}
\end{figure}

Choosing $t_0=T$ generates the profile of $\mu_D$, $t_0=0$ $\Gamma_L$, and $t_0=2T$ yields $\Gamma_R$.
The resulting profiles can be found in Fig. \ref{fig:app_profiles}.

The parameters for the time evolution need to obey the adiabaticity constraint $T\sim h/\Gamma_0$ of the protocol.
To find parameters for the time evolution that both, obey the adiabaticity constraint and deliver unit fidelities, we construct the full Hamiltonian as given in Eq.~\eqref{eq:H_total}.
We fix all the parameters at their optimal point ($t_L=\Delta_L=t_R=\Delta_R=5\Gamma_0$, and $\varphi=\pi$).
We choose the stepsize of the time discretization to $\Delta t=0.2 h/\Gamma$ corresponding to $\sim 0.1ns$ given typical coupling strengths of $\Gamma_0 \sim 10\mu eV$.
We then optimize numerically for $T$ and the ramping time $t_{ramp}$ by demanding that $P_{|oo\rangle}(3T)=1/2$ and $P_{|oo\rangle}(6T)$=1.
This results in $(T_{opt}, t_{ramp, opt})=(200.54, 21.21) \hbar/\Gamma_0$ which is our default parameter choice unless specified otherwise. 
When changing $T$ away from the optimized value, we have to adjust the ramping time accordingly. 
This we do by letting $t_{ramp}\rightarrow T t_{ramp,opt}/T_{opt}$ such as to coincide with the optimal choice when $T=T_{opt}$.

\section{Effective odd parity Hamiltonian and excitation minimum} \label{sec:app_coulomb}

\subsection{Effective Hamiltonian in odd and even parity sectors}
As discussed in the main text, it suffices to only consider a single PMM coupled to the ancillary dot to understand the physics relevant for the Coulomb repulsion.
The Hamiltonian of this system reads

\begin{align} \label{eq:system_hamiltonian}
& H_{LD} = H_{K,L} + H_{\mathrm{tunn}, L} + H_{C,L} + H_D, \nn
& H_{K,L}= \mu_{L1} n_{L1} + \mu_{L2}n_{L2} + t c\dg_{L2} c_{L1} + \Delta c_{L2} c_{L1} + h.c., \nn
& H_D = \mu_D n_D, \nn
& H_{\rm{tunn},L} = \Gamma_L \left(c\dg_D c_{L2} + c_{L2}\dg c_D\right), \nn
& H_{C,L} = U_L n_D n_{L2}.
\end{align}

Since the system preserves the total fermionic parity we can separate the Hilbert space into the even and odd total parity subspaces.
In the even parity sector consisting of the basis $|e_L,0_D\rangle, |o_L,1_D\rangle, |e_L^\prime,0_D\rangle, |o^\prime_L, 1_D\rangle$ we find Eq.~\eqref{eq:h_ld}, i.e.
\begin{align} \label{eq:h_ld_even_app}
    H_{LD}^{(even)} = \bpm
    0 & \frac{\Gamma_L}{2} & 0 & \frac{\Gamma_L}{2} \\
    \frac{\Gamma_L}{2} & \mu_D + \frac{U_L}{2} & \frac{\Gamma_L}{2} & -\frac{U_L}{2} \\
    0 & \frac{\Gamma_L}{2} & 2\Delta_L  & \frac{\Gamma_L}{2} \\
    \frac{\Gamma_L}{2} & -\frac{U_L}{2} & \frac{\Gamma_L}{2} & 2\Delta_L + \mu_D + \frac{U_L}{2}
    \epm.
\end{align}
In the odd parity basis, consisting of $|e_L, 1_D\rangle,|o_L, 0_D\rangle,|e^\prime_L, 1_D\rangle,|o^\prime_L, 0_D\rangle$, we find through an analogous calculation the Hamiltonian

\begin{align} \label{eq:h_ld_odd_app}
    H_{LD}^{(odd)} = \begin{pmatrix}
        \frac{U_L}{2}+\mu_D & \frac{\Gamma_L}{2}& \frac{U_L}{2}&-\frac{\Gamma_L}{2} \\
        \frac{\Gamma_L}{2} & 0 & -\frac{\Gamma_L}{2} & 0 \\
        \frac{U_L}{2} & -\frac{\Gamma_L}{2} & \frac{U_L}{2}+2\Delta_L+\mu_D & \frac{\Gamma_L}{2} \\
        -\frac{\Gamma_L}{2} & 0 & \frac{\Gamma_L}{2} & 2\Delta_L
    \end{pmatrix}, 
\end{align}
Due to the aforementioned parity conservation the spectrum will be strongly degenerate between parity sectors.

\subsection{Excitation gap minimum}

The excitation gap can be found through Eq.~\eqref{eq:H_eff} from the main text.
We find the eigenvalues of the effective Hamiltonian to be

\begin{align}
    \epsilon_\pm &= \frac{1}{2}\left(\mu_D+\frac{U_L}{2}+\Delta_L-\lambda\right)\pm\sqrt{\frac{\Gamma_L^2}{4}(a+b)^2+\frac{1}{4}\left(\mu_D+\frac{U_L}{2}+\Delta_L-\lambda\right)^2}.
\end{align}

The excitation gap is the difference of these two eigenvalues given as

\begin{align} \label{eq:exc_gap}
\Delta\epsilon = \sqrt{\Gamma_L^2(a+b)^2+\left(\mu_D+\frac{U_L}{2}+\Delta_L-\lambda\right)^2}
\end{align}

Inspecting Eq.~\eqref{eq:exc_gap}, it becomes apparent that the excitation gap becomes minimal exactly for the predicted value of $\mu_D=\mu_D^*$ as given in Eq.~\eqref{eq:mu_Dstar}.
We find the value of the excitation gap to be

\begin{align}
    \Delta\epsilon_{\mathrm{min}}=\Gamma_L\frac{\sqrt{\sqrt{4\Delta_L^2+U_L^2}+2\Delta_L}+\sqrt{\sqrt{4\Delta_L^2+U_L^2}-2\Delta_L}}{\sqrt{2}(4\Delta_L^2+U_L^2)^{1/4}}
\end{align}

for a single Kitaev chain attached to the ancillary dot.

\section{Calculation of the non-Abelian Berry phase}\label{sec:berry_phase}

In this section, we show the details for calculating the non-Abelian Berry's phase in the presence of residual couplings.
The calculation is similar in spirit to Ref.~\cite{venHeck2012Coulomb}, but we generalize it to asymmetric couplings.
The Hamiltonian is given by 
\begin{align}
H = \sum^3_{k=1} \Delta_k i \gamma_0 \gamma_k,
\end{align}
which involves four Majoranas in total.
We thus define fermionic operators as below
\begin{align}
 f_1 = (\gamma_1 + i \gamma_2)/2, \quad f_2 = (\gamma_0 + i \gamma_3)/2,
\end{align}
so that we can further define the four-dimensional Fock space as
\begin{align}
&\ket{00}, \nn
&\ket{10} = f\dg_1 \ket{00}, \nn
&\ket{01} = f\dg_2 \ket{00}, \nn
&\ket{11} = f\dg_1 f\dg_2 \ket{00}.
\end{align}
As such, the Hamiltonian can now be written as
\begin{align}
H = 
\bpm
\ket{00} \\
\ket{11} \\
\ket{10} \\
\ket{01}
\epm^T
\bpm
-\Delta_3 & i \Delta_1 + \Delta_2 & 0 & 0 \\
-i \Delta_1 + \Delta_2 & \Delta_3 & 0 & 0 \\
0 & 0 & -\Delta_3 & -i \Delta_1+\Delta_2 \\
0 & 0 & i \Delta_1 +\Delta_2  & \Delta_3 
\epm
\bpm
\bra{00}\\
\bra{11}\\
\bra{10}\\
\bra{01}
\epm.
\end{align}
We note that here the even- and odd-parity subspaces are block diagonalized due to fermion parity conservation.
The dimension of each subspace is two.

We first focus on the even-parity subspace, where the ground-state energy is 
\begin{align}
E_{e, gs} = -\varepsilon = - \sqrt{\Delta^2_1 + \Delta^2_2 +\Delta^2_3},
\end{align}
and the wavefunction is 
\begin{align}
\ket{e} = \sqrt{\frac{\varepsilon - \Delta_3}{2\varepsilon }}
\bpm
-i \frac{\Delta_3 + \varepsilon}{\Delta_1 + i \Delta_2} \\
1
\epm.
\end{align}
Using sympy, we obtain that 
\begin{align}
&A_{e,1} = \langle e | \frac{d}{d\Delta_1} | e \rangle = \frac{\Delta_2}{\Delta^2_1 + \Delta^2_2} \frac{i( \varepsilon + \Delta_3 )}{2\varepsilon }, \nn
&A_{e,2} = \langle e | \frac{d}{d\Delta_2} | e \rangle = \frac{-\Delta_1}{\Delta^2_1 + \Delta^2_2} \frac{i( \varepsilon + \Delta_3 )}{2\varepsilon }, \nn
&A_{e,3} = 0.
\label{eq:A}
\end{align}

Using the same calculation method, we find that
\begin{align}
E_{o, gs} = -\varepsilon = - \sqrt{\Delta^2_1 + \Delta^2_2 +\Delta^2_3},
\end{align}
and the wavefunction is 
\begin{align}
\ket{o} = \sqrt{\frac{\varepsilon - \Delta_3}{2\varepsilon }}
\bpm
i \frac{\Delta_3 + \varepsilon}{\Delta_1 - i \Delta_2} \\
1
\epm.
\end{align}
Note that the odd-parity wavefunction is different from the even one by $i \to -i$. Thus the signs of the Berry connections are simply reversed, i.e.
\begin{align}
\vec{A}_{o} = - \vec{A}_{e}.
\end{align}
Due to parity conservation, the matrix elements between even- and odd-parity states are zero.

The unitary evolution for the Majorana braiding is defined as 
\begin{align}
U = \exp \left( - \oint_c \sum_k A_k d \Delta_k  \right).
\label{eq:U}
\end{align} 
Here the Berry connection shown in Eq.~\eqref{eq:A} is singular when $\Delta_1=\Delta_2=0$ because of the presence of a term $\Delta_1/(\Delta^2_1 + \Delta^2_2)$.
To avoid the singular points, we assume that the couplings have some residual amplitudes even when they are ``switched off'', i.e., $\eta_k \leq \Delta_k \leq \Gamma$.
Here we first assume that the maximal strengths of all $\Delta_k$'s are assumed to be the same and equal to $\Gamma$ to simplify the calculation.
In particular, in the three-step braid operation, we assume six contours in the parameter path as below
\begin{align}
    & C1: (\eta_1, \eta_2, \Gamma) \to (\Gamma, \eta_2, \Gamma), \nn
    & C2: (\Gamma, \eta_2, \Gamma) \to (\Gamma, \eta_2, \eta_3), \nn
    & C3: (\Gamma, \eta_2, \eta_3) \to (\Gamma, \Gamma, \eta_3), \nn
    & C4: (\Gamma, \Gamma, \eta_3) \to (\eta_1, \Gamma, \eta_3), \nn
    & C5: (\eta_1, \Gamma, \eta_3) \to (\eta_1, \Gamma, \Gamma), \nn
    & C6: (\eta_1, \Gamma, \Gamma) \to (\eta_1, \eta_2, \Gamma), \nn
\end{align} 
where each bracket denotes $(\Delta_1, \Delta_2, \Delta_3)$.
We note that only 4 out of the 6 contours contribute to the Berry's phase.
We name them as $C1 \to I_1, C4 \to I_2, C3 \to I_3, C6 \to I_4$.
\begin{align}
& - \oint_c \sum_k A_k d \Delta_k = \left(\frac{-i}{2} \right) I =\left(\frac{-i}{2} \right) (I_1 +  I_2 +  I_3 +  I_4),
\end{align} 
In particular
\begin{align}
I_1 = \int^{\Gamma}_{\eta_1} A_1(\Delta_1, \eta_2, \Gamma) d \Delta_1  = \int^{\Gamma}_{\eta_1} d\Delta_1 \frac{\eta_2}{\Delta^2_1 + \eta^2_2} \left( 1 + \frac{\Gamma}{\sqrt{\Delta^2_1 + \Gamma^2}}  \right).
\end{align}
Here the first integral is
\begin{align}
I_{11}=\int^{\Gamma}_{\eta_1} d\Delta_1 \frac{\eta_2}{\Delta^2_1 + \eta^2_2} = \int^{b}_{a} dx \frac{1}{x^2 + 1}
= \atan(b) - \atan(a),
\end{align}
where $b=\Gamma/\eta_2$ and $a=\eta_1/\eta_2$.
The second part is
\begin{align}
I_{12}=\int^{\Gamma}_{\eta_1} d\Delta_1 \frac{\eta_2}{\Delta^2_1 + \eta^2_2} \frac{\Gamma}{\sqrt{\Delta^2_1 + \Gamma^2}} = \int^b_a  dx \frac{1}{x^2 + 1} \frac{b}{\sqrt{x^2+b^2}}.
\end{align}
We obtain the result for the indefinite integral as given below
\begin{align}
\int  dx \frac{1}{x^2 + 1} \frac{b}{\sqrt{x^2+b^2}} = \frac{b}{\sqrt{1-b^2}} \tanh^{-1} \left(  \frac{\sqrt{1-b^2} }{\sqrt{b^2+x^2}} x \right)
\approx \atan \left(  \frac{bx }{\sqrt{b^2+x^2}}  \right),
\end{align}
where we assume $b \gg 1$ and use the relation of $\tanh^{-1}(ib) = i \atan(b)$.
Therefore we have
\begin{align}
I_{12} = \atan(\frac{b}{\sqrt{2}}) - \atan(a) \approx \frac{\pi}{2} - \frac{\sqrt{2}}{b} - \atan(a),
\end{align}
where we consider $a \sim O(1) \ll b$ and use the identity of $\atan(x) = \pi/2 - \atan(1/x)$.
We thus have
\begin{align}
I_1 =  \atan(\Gamma/\eta_2) - 2\atan(\eta_1/\eta_2) + \frac{\pi}{2} - \frac{\sqrt{2} \eta_2}{\Gamma}.
\end{align}
Next, the second integral is 
\begin{align}
I_2 &= \int^{\eta_1}_{\Gamma} A_1(\Delta_1, \Gamma, \eta_3) d \Delta_1  = -\int^{\Gamma}_{\eta_1} d\Delta_1 \frac{\Gamma}{\Delta^2_1 + \Gamma^2} \left( 1 + \frac{\eta_3}{\sqrt{\Delta^2_1 + \Gamma^2}}  \right) \nn
&= -\left[ \atan(1) - \atan(\eta_1/\Gamma) + \frac{\eta_3}{\sqrt{2} \Gamma} \right].
\end{align}
And the third integral is
\begin{align}
I_3 &= \int^{\Gamma}_{\eta_2} A_2(\Gamma, \Delta_2, \eta_3) d \Delta_2  
= \int^{\Gamma}_{\eta_2} d\Delta_2 \frac{-\Gamma}{\Delta^2_2 + \Gamma^2} \left(1 +  \frac{\eta_3}{\sqrt{\Delta^2_2 + \Gamma^2}} \right) \nn
&= -\left[ \atan(1) - \atan(\eta_2/\Gamma) + \frac{\eta_3}{\sqrt{2} \Gamma} \right].
\end{align}
Lastly, the fourth integral is
\begin{align}
I_4 &= \int^{\eta_2}_{\Gamma} A_2(\eta_1, \Delta_2, \Gamma) d \Delta_2 \nn
&= \atan(\Gamma/\eta_1) + \frac{\pi}{2} - \frac{\sqrt{2} \eta_1}{\Gamma} - 2\atan(\eta_2/\eta_1).
\end{align}
After summing them up, we obtain 
\begin{align}
I = \sum^4_{i=1} I_i = \frac{\pi}{2} - \frac{\sqrt{2}}{\Gamma}( \eta_1 + \eta_2 + \eta_3 ),
\end{align}
giving the unitary evolution matrix as below
\begin{align}
U = \exp \left( \frac{-i}{2} I \sigma_z \right) = \exp \left( -i \left( \frac{\pi}{4} - \epsilon \right) \sigma_z \right),
\end{align}
where 
\begin{align}
\epsilon = \frac{(\eta_1 + \eta_2 + \eta_3)}{ \sqrt{2}\Gamma}  + O(\eta_i/\Gamma)^2.
\end{align}
In the notation for braiding Majoranas in Kitaev chains, it becomes
\begin{align}
U_{\rm{braid}} = \exp \left( -i \left( \frac{\pi}{2} - \epsilon \right) \gamma_2\gamma_3 \right),
\end{align}
where 
\begin{align}
\epsilon &= \frac{(\Gamma_{L,min} + \Gamma_{R,min} + \mu_{D,min})}{ \sqrt{2}\Gamma}  + O(\eta_i/\Gamma)^2 
\end{align}
A similar calculation can be performed assuming $\Delta_{1,max}=\Delta_{2,max}\ll \Delta_{3,max}$.
It yields 
\begin{align}
    \epsilon = \frac{1}{\sqrt{2}}\left( \frac{\Gamma_{L,\text{min}}}{\Gamma} + \frac{\Gamma_{R,\text{min}}}{\Gamma} + \frac{\sqrt{2}\mu_{D,\text{min}}}{\Gamma} \right) + O(\eta_i/\Gamma, \Gamma/\mu_{D,max}).
\end{align}

\end{document}